# Stability of nucleus-acoustic waves in completely degenerate white dwarf cores and their nearly degenerate ambience

Sayanti Dasgupta[a], Ahmed Atteya[b] and Pralay Kumar Karmakar[a,*]
[a]Department of Physics, Tezpur University, Napaam-784028, Tezpur, Assam, India
[b]Department of Physics, Faculty of Science, Alexandria University, P.O. 21511, Alexandria, Egypt
[*]Corresponding E-mail: pkk@tezu.ernet.in

**ABSTRACT**
We analyze the propagatory nucleus-acoustic wave (NAW) modes excitable in the completely degenerate (CD) core and in its nearly degenerate (ND) ambience of the ONe and CO white dwarfs (WDs). It is based on three-component spherical hydrodynamic quantum plasma consisting of tiny non-thermal quantum electrons, classical thermal light nuclear species (LNS), and classical thermal heavy nuclear species (HNS). The inner concentric layer-wise electronic pressures are judiciously modelled. The electronic energy distribution governed by the Fermi-Dirac (FD) thermostatistical distribution law involves both the thermodynamical temperature and chemical potential. Our exploration emphasizes on the transition state between the thermodynamical temperature and the Fermi temperature for the borderline regions of intermediate degeneracy. A normal spherical mode analysis procedurally yields a sextic generalized linear dispersion relation highlighting the plasma multiparametric dependency of the NAW-features. A numerical illustrative platform is constructed to investigate the full NAW propagatory and dispersive behaviours. We demonstrate that the NAW in ONe (CO) WDs exhibits sensible growth characteristics at near the transcritical (supercritical) wave zone. The temperature-sensitivity of the NAW-growth is more (less) prominent in ONe (CO) WDs. It could be hopefully useful to see the internal structure of compact astroobjects from the asteroseismic probe-perspective of collective quantum interaction processes.

## 1. Introduction

The area of quantum plasmas is one of the most sought after research fields owing to its wide spectrum of promising applications ranging from the nanoscales to the astrocosmic scales of space and time. It has widespread scope in the field of modern technological advancements, such as metallic nanoparticles, thin metal films, nanotubes, quantum X-ray free electron lasers, and so forth [1]. In the astrophysical context, they exist in interiors of white dwarfs (WDs), magnetars, jovian planets, and so on [1, 2]. WDs are the end products of stellar evolution for most of the low and medium mass main sequence stars [2]. After the hydrogen fusion ends, the core temperature is sufficient to fuse helium (He), leading to formation of carbon (C) and oxygen (O). The outer layers expand and cool, thereby forming a red giant. The star then sheds its outer layers, forming a gaseous shell (planetary nebula) around the core [3]. This remnant core with no fuel left to counter the inward self-gravity action forms the WD. In most cases, the core is made up of C and O, forming a CO WD [3-5]. For stars having masses ($M$) in the range $8M_\Theta \lesssim M \lesssim 11M_\Theta$ (where, $M_\Theta = 1.98 \times 10^{30}$ kg is the solar mass), the temperature is sufficient to fuse C, but not neon (Ne), leading to the formation of ONeMg cores of WDs [6, 7].

When we consider the electrons present in the WDs, the consideration of quantum-mechanical interactions, like the exchange and correlation interactions become of utmost importance. It is noteworthy to mention that these interactions do not have any classical analogs, and therefore are purely quantum-mechanical. The exchange energy is a direct consequence of the Pauli exclusion principle [8, 9]. As a result, electrons with same spin tend to repel each other. In other words, electrons having anti-symmetric spins already repel each other, thereby reducing the Coulomb repulsion that would otherwise exist between them in absence of the same spin condition [8, 9]. Correlation interaction gives a metric to determine how much the mobility of one electron is influenced by the presence of the surrounding electrons [9-11]. Mathematically, correlation energy is the difference between the total electronic energy and the energy obtained from Hartree-Fock approximation after simplifying a many-body quantum-mechanical system into an equivalent single one via the Slater determinant [11].

It is noteworthy that the impact of exchange and correlation interactions have been studied in quite a few number of quantum systems in the past [12-20]. The dispersive properties of bounded quantum plasma with the electron exchange-correlation effects in nano-cylindrical waveguides have been studied analytically and numerically [12]. The surface plasmon oscillations in semi-bounded quantum plasma (metallic plasmas and laser produced solid density plasmas) have also been studied [13]. The wake potential in the presence of upper hybrid waves in magnetized semiconductor quantum plasmas have been studied under the influence of quantum effects like exchange and correlation [14]. In the field of semiconductor quantum plasmas, the propagatory features of extraordinary electromagnetic (X-EM) waves in magnetized electron-positron plasmas under the influence of the exchange-correlation and Bohm potential have also been thoroughly investigated [15]. The collective influence of the exchange-correlation potential, Fermi velocity, and Bohm force on the lower hybrid waves responsible for electron acceleration in plasma heating mechanism has also been investigated [16]. Again, the oscillatory wake potential of a mobile test charge has been studied in magnetized quantum dusty plasma in the presence of several important quantum effects, like tunneling, degeneracy, and exchange-correlation effects [17]. The influence of the electron exchange-correlation field on magneto-acoustic-gravitational instability, useful in the context of self-gravitating magnetoplasma systems, has also been investigated [18]. Linear and non-linear drift ion-acoustic waves have been studied under the conjoint quantum influences like electron tunneling and exchange-correlation [19]. The propagation characteristics of lower hybrid waves in electron-positron degenerate plasma in presence of electron exchange-correlation have been thoroughly investigated in the non-relativistic, relativistic, and ultra-relativistic regimes [20]. High-frequency surface waves excited in dense astrophysical plasmas have been explored, where the quantum electrons were subject to Bohm potential, spin magnetization energy, relativistic degenerate pressure, and exchange-correlation effects [21]. The effects of exchange and correlation have been theoretically investigated on filamentation instability of high-density current-driven plasma [22]. The role of electron exchange correlation and positron exchange correlation on linear and non-linear ion-acoustic waves has also been analytically investigated [23]. Very recently, the oblique propagation of electrostatic waves in degenerate magnetized quantum plasmas has been theoretically explored [24]. Non-linear inertial Alfven waves have also been studied under the conjoint influence of electron exchange-correlation and spin magnetization in electron-ion quantum plasma [25]. Thus, we see that the electron exchange and correlation interactions play a major and important role in different types of quantum plasmas having wide range applicability.

In our semi-analytic theoretic study, we investigate the excitation, propagation, and dispersion characteristics of the nucleus-acoustic waves (NAWs) in the completely degenerate (CD) ONe and CO cores, and in the nearly degenerate (ND) surrounding region around the core, depicting the transition between thermodynamic temperature, $T$, and Fermi temperature, $T_F$, explicitly for both the considered WDs. The system is composed of three constitutive species, namely the electrons, the light nuclear species (LNS) ($^{12}_{6}C$ for CO WD and $^{16}_{8}O$ for ONeMg WD) and the heavy nuclear species (HNS) ($^{16}_{8}O$ for CO WD and $^{20}_{10}Ne$ for ONeMg WD).

The electrons are governed by an equation of state (EoS) that was developed exclusively for the study of the WDs [8, 9]. The above mentioned EoS takes into account the contribution due to the electronic pressure (degenerate Fermi pressure for the CD core and the ND pressure for the surrounding transition ambience around the core), pressure due to the interaction of the electrons with other electrons and surrounding nuclei, exchange interaction, and correlation interaction, explicitly. The constitutive LNS and HNS are governed classically by an appropriate EoS taking into account their thermal pressures.

The quantum electrons, governed by the Fermi-Dirac (FD) statistical distribution law, are characterized by two important parameters: temperature, $T$ and chemical potential, $\mu$ [26-29]. In addition to the CD pressure in the core of the considered WDs, our study also emphasizes on the ND pressure in the borderline region (with an intermediate degree of degeneracy) around the core which is neither strongly degenerate, nor strongly non-degenerate. This is achieved by means of a temperature degeneracy parameter, expressed with a usual symbolism (unfolded later in the text), as $G'_e = Li_{5/2}(-\xi)/Li_{3/2}(-\xi)$ [26-29]. The degeneracy of the system is described with the help of $\xi = e^{\beta'\mu}$ (thermodynamic beta, $\beta' = 1/k_B T$), a function of $\mu$ and $T$ [26-29]. A large number of observational evidences and stellar evolutionary models have been reported in favour of the ONeMg WDs [6, 7, 30-33]. Recent model calculations have established O and Ne as the most significant components of the dwarf core, thereby reducing the Mg abundance [7]. Likewise, several models and observational findings have also been reported in favour of the CO WDs [4, 5, 34]. In fact, very recently, IRAS00500+6713, an object having super-Chandrasekhar mass, has been observationally reported, which is believed to be a merger product of a ONe and CO WD [35]. Thus, we see that there are quite a few numbers of studies dealing with observational and astrophysical aspects of CO and ONe WDs. A large number of semi-analytic investigations on NAWs have also been reported [36-39]. However, the study of plasma wave excitation and propagation in these WDs explicitly (CD core and ND transition region around the core of ONeMg and CO WDs), has still been lying as an open problem that is yet to be well-addressed. In our proposed model analysis, we investigate the same in a semi-classical and semi-analytic approach with all the said relevant realistic key factors taken into account.

Apart from the general introduction and brief specifications of the proposed model in Section 1, the physical model and mathematical formulation is given in Section 2. Section 3 deals with the perturbation scheme and the normal spherical mode analysis resulting in the generalized dispersion relation. Our results and discussions are given in Section 4. Finally, the concluding remarks, along with a brief highlight on realistic applications and implications, are summarily reported in Section 5.

## 2. Model formulation

We consider a theoretic quantum hydrodynamic model in a spherically symmetric geometrical construct to study the degeneracy-dependent radial WD-core behaviour of the NAW stability, dispersion, and propagation. The model consists of three constitutive species; namely, quantum electrons, classical LNS, and classical HNS. The EoS of the quantum electrons takes into consideration the contributions due to the electronic pressure resulting from temperature degeneracy (both CD pressure in the core and ND pressure in the transition region around the core demarcating transition between $T$ and $T_F$) [26], interaction of electrons with other electrons and surrounding nuclei, exchange interaction and correlation interaction, explicitly [8, 9]. The classical LNS and HNS are governed by an appropriate EoS taking into account their respective thermal pressures. The dynamics of all the three constitutive species are governed by the flux conservation continuity equation, the force-balancing momentum equation, and their respective proper forms of the EoS. The model closure is finally obtained by means of the electrostatic and self-gravitational Poisson equations.

The respective equations governing the electronic dynamics with all generic notations [36, 37] are accordingly cast as

$$\partial_t n_e + r^{-2}\partial_r(r^2 n_e u_e) = 0, \tag{1}$$

$$e\partial_r\phi - n_e^{-1}\partial_r P_e - \hbar^2(2m_e)^{-1}\partial_r\left(n_e^{-\frac{1}{2}}\left[r^{-2}\partial_r\left\{r^2\partial_r\left(n_e^{-\frac{1}{2}}\right)\right\}\right]\right) = 0, \tag{2}$$

$$P_e = P_{el} - P_{ie} - P_{exc} - P_{corr}. \tag{3}$$

Similarly, the basic equations governing the dynamics of the classical particles (with $\alpha = l$ for LNS and $\alpha = h$ for HNS) are given as

$$\partial_t n_\alpha + r^{-2}\partial_r(r^2 n_\alpha u_\alpha) = 0, \tag{4}$$

$$\partial_t u_\alpha + (eZ_\alpha m_\alpha^{-1})\partial_r\phi + \partial_r\psi + (m_\alpha n_\alpha)^{-1}\partial_r(P_\alpha) = 0, \tag{5}$$

$$P_\alpha = n_\alpha k_B T. \tag{6}$$

The system closing electrostatic Poisson equation taking into account the electrostatic interactions of all the three species is given as

$$r^{-2}\partial_r(r^2\partial_r\phi) = (e\varepsilon_0^{-1})(n_e - Z_l n_l - Z_h n_h). \tag{7}$$

Likewise, the self-gravitational Poisson equation is written as

$$r^{-2}\partial_r(r^2\partial_r\psi) = (4\pi G)(\Delta\rho_l + \Delta\rho_h). \tag{8}$$

All the symbols and values used here are quite in a customary form [26-29, 36, 37]. The notations $n_e$ and $u_e$ denote the number density and flow speed of the electronic species. $e = 1.6\times10^{-19}$ C denotes the electronic charge. $m_e = 9.1\times10^{31}$ kg is the electronic mass. $\hbar = h/2\pi \sim 10^{-34}$ J s is the reduced Planck constant. $P_e$ denotes the effective electronic pressure, which is composed of the pressure due to the temperature degeneracy, interaction with surrounding nucleons, exchange and correlation interaction. $P_{el} = G_e' n_e/\beta'$ stands for the electronic pressure due to the temperature degeneracy.

We employ an explicit function describing the temperature degeneracy parameter defined for the transition between $T$ and $T_F$ in generic notations [26-29] as

$$G_e' = Li_{5/2}(-\xi)/Li_{3/2}(-\xi). \tag{9}$$

Here, $Li_p(-\xi)$ is the polylogarithmic function with index $p$ with $\xi(\mu,T) = e^{\beta'\mu} = e^{\mu/k_B T}$ [26-29]. The general form of $Li_p(-\xi)$ signifying temperature degeneracy effects [26-29] for $p > 0$ is

$$Li_p(-\xi) = -(\Gamma(p))^{-1} \int_0^\infty t^{p-1}\left(e^t \xi^{-1} + 1\right)^{-1} dt, \tag{10}$$

where, $\Gamma(p) = \int_0^\infty x^{p-1} e^{-x} dx$ is the gamma function.

For the CD limit ($\xi \to \infty$), we get

$$G'_e = 2(5\delta)^{-1}, \text{ where, } \delta = T/T_F, \tag{11}$$

and for the ND limit ($\xi \gg 1$), we get

$$G'_e = 2(5\delta)^{-1}\left\{1 - (\pi\delta)^2(12)^{-1}\right\}. \tag{12}$$

$k_B = 1.38 \times 10^{-23}$ J K$^{-1}$ is the Boltzmann constant signifying energy-temperature correlationship. $P_{ie} = 0.48 e^2 n_e^{4/3}$ gives the resultant pressure due to the interaction of electrons with other neighbouring electrons and surrounding nuclei [8, 9]. $P_{exc} = 0.25 e^2 n_e^{4/3}$ stands for the pressure due to the electronic exchange interactions [8, 9]. $P_{corr} = 0.0104 e^2 n_e/a_0$ gives the pressure due to the electronic correlation interactions [8, 9]. $a_0 = 5.29 \times 10^{-11}$ m is the Bohr unit (Bohr radius). It is noted that Equation (3) is developed specially for modelling the WDs interiors [8].

As already mentioned above, $\alpha = l$ for the LNS ($^{12}_{6}C$ for CO WD and $^{16}_{8}O$ for ONeMg WD), and $\alpha = h$ for the HNS ($^{16}_{8}O$ for CO WD and $^{20}_{10}Ne$ for ONeMg WD) in Equations (4)-(6). $n_\alpha$ and $u_\alpha$ stand for the number density and flow velocity. $Z_\alpha$ denotes their charge states. Likewise, $P_\alpha$ stands for the thermal pressure. $\phi$ and $\psi$ give the electrostatic and gravitational potentials, respectively. $\varepsilon_0 = 8.85 \times 10^{-12}$ F m$^{-1}$ denotes the permittivity of the plasma medium. In Equation (8), $\Delta\rho_l = \rho_l - \rho_{l0} = m_l(n_l - n_{l0})$ and $\Delta\rho_h = \rho_h - \rho_{h0} = m_h(n_h - n_{h0})$ to model the Jeans swindle. $G = 6.67 \times 10^{-11}$ N m$^2$ kg$^{-2}$ is the universal gravitational coupling constant.

A number of points regarding the above equations are noteworthy. Equation (1) is the equation of continuity depicting the conservation of flux of the electronic species. Equation (2) is the force-balancing momentum equation, where the forces due to the electrostatic potential (1$^{st}$ term), electronic pressure (2$^{nd}$ term), and Bohm potential (3$^{rd}$ term) exactly balance each other. Equation (3) is the EoS for the electronic species, taking into account the pressures due to temperature degeneracy (1$^{st}$ term), interaction of electrons with other electrons and surrounding nuclei (2$^{nd}$ term), exchange interaction (3$^{rd}$ term), and correlation interaction (4$^{th}$ term). Equation (4) denotes the equation of continuity for the classical species (LNS+HNS). Likewise, Equation (5) is the analog of Equation (2), but for classical LNS and HNS, where the forces by virtue of their motion (1$^{st}$ term), electrostatic potential (2$^{nd}$ term), gravitational potential (3$^{rd}$ term), and thermal pressure (4$^{th}$ term) are exactly balanced by each other. Equation (6) is the EoS taking into account their thermal pressure. Equations (7)-(8) give the closure of the system in terms of the electrostatic and self-gravitational Poisson equations, respectively.

For a scale-invariant analysis, we employ a standard astronomical normalization scheme [36-38]. The dimensionless set of the basic governing equations are now cast as

$$\partial_\tau N_e + R^{-2}\partial_R(R^2 N_e M_e) = 0, \tag{13}$$

$$N_e \partial_R \Phi - \left( p_{dp}^* N_e^{\frac{2}{3}} \partial_R N_e - p_{ie}^* N_e^{\frac{1}{3}} \partial_R N_e - p_{exc}^* N_e^{\frac{1}{3}} \partial_R N_e - p_{corr}^* \partial_R N_e \right)$$
$$- 4^{-1} H'^2 M_{Fe}^2 \left( \partial_R^3 N_e + 2R^{-1}\partial_R^2 N_e - 2R^{-2}\partial_R N_e \right) = 0, \qquad \text{(for CD case)} \tag{14.1}$$

$$N_e \partial_R \Phi - \left( G'_e T^* \partial_R N_e - p_{ie}^* N_e^{\frac{1}{3}} \partial_R N_e - p_{exc}^* N_e^{\frac{1}{3}} \partial_R N_e - p_{corr}^* \partial_R N_e \right)$$
$$- 4^{-1} H'^2 M_{Fe}^2 \left( \partial_R^3 N_e + 2R^{-1}\partial_R^2 N_e - 2R^{-2}\partial_R N_e \right) = 0. \qquad \text{(for ND case)} \tag{14.2}$$

The dynamics of the classical LNS in normalized form are given as

$$\partial_\tau N_l + R^{-2}\partial_R(R^2 N_l M_l) = 0, \tag{15}$$
$$N_l \partial_\tau M_l + N_l \partial_R \Phi + N_l \partial_R \Psi + A_{el}\partial_R(N_l T^*) = 0. \tag{16}$$

Analogously, the dynamics of the HNS in normalized form are given as

$$\partial_\tau N_h + R^{-2}\partial_R(R^2 N_h M_h) = 0, \tag{17}$$
$$N_h \partial_\tau M_h + N_h \beta \partial_R \Phi + N_h \partial_R \Psi + A_{eh}\partial_R(N_h T^*) = 0. \tag{18}$$

The system closing electrostatic and self-gravitational Poisson equations in dimensionless forms are respectively given as

$$R^{-2}\partial_R(R^2 \partial_R \Phi) = N_e(1+\mu') - N_l - \mu' N_h, \tag{19}$$
$$R^{-2}\partial_R(R^2 \partial_R \Psi) = \sigma\{(N_l - 1) + \mu'\beta^{-1}(N_h - 1)\}. \tag{20}$$

In the above Equations (13)-(20), the spatial coordinate is normalized as $R = r/\lambda_{Dl}$; where, $\lambda_{Dl} = (m_e c^2 \varepsilon_0 / n_{l0} Z_l e^2)^{1/2}$ is the light nuclear Debye length. The temporal coordinate is normalized as $\tau = t/\omega_{pl}^{-1}$; where, $\omega_{pl} = (n_{l0} Z_l^2 e^2 / m_l \varepsilon_0)^{1/2}$ is the light nuclear plasma oscillation frequency. Normalized population density of the constitutive particles is given as $N_s = n_s/n_{s0}$; where, $n_{s0}$ is the equilibrium population density, $s$ being $e, l, h$ for the electronic species, LNS, and HNS respectively. Normalized form of flow velocity is given by $M_s = u_s/C_l$; where, $C_l = (Z_l m_e c^2/m_l)^{1/2}$ is the light nuclear transit speed. The normalized CD pressure coefficient is given as $p_{dp}^* = p_{dp}/m_e c^2$; where, $p_{dp} = 1.51 e^2 n_{e0}^{2/3}$ is the CD pressure coefficient. The normalized pressure coefficient due to interaction of electrons with other electrons and nuclei is given as $p_{ie}^* = p_{ie}/m_e c^2$; where, $p_{ie} = 0.64 e^2 n_{e0}^{1/3}$ is the unnormalized pressure coefficient due to similar electronic interactions. The normalized pressure coefficient due to exchange interaction of the electrons is given as $p_{exc}^* = p_{exc}/m_e c^2$; where, $p_{exc} = 0.33 e^2 n_{e0}^{1/3}$ is the unnormalized pressure coefficient resulting from exchange interaction. $p_{corr}^* = p_{corr}/m_e c^2$ gives the normalized pressure coefficient due to correlation interaction of the electrons. Here, $p_{corr} = 0.0104 e^2/a_0$ is the unnormalized pressure coefficient due to correlation interaction of the electrons. $T^* = T k_B/m_e c^2$ denotes the normalized temperature. $H' = \hbar \omega_{pl}/m_e v_{Fe}^2$ denotes the quantum parameter signifying the ratio of the plasmon energy associated with the LNS to that of the Fermi energy associated with the electronic species. The Fermi Mach number is given as $M_{Fe} = v_{Fe}^2/C_l c$; where, $v_{Fe}$ is the Fermi velocity. $\beta = Z_h m_l/Z_l m_h$ is the relative nuclear charge-to-mass coupling parameter. $\mu' = Z_h n_{h0}/Z_l n_{l0}$ is the ratio of the charge densities of the heavy-to-LNS. $A_{el} = m_e c^2/m_l C_l^2$ stands for the ratio of the relativistic electronic energy to that of the LNS energy. $A_{eh} = m_e c^2/m_h C_l^2$ is the analogous term for the HNS. The ratio of the square of the Jeans frequency to light nuclear plasma oscillation frequency is given as $\sigma = \omega_{Jl}^2/\omega_{pl}^2$; where, $\omega_{Jl} = (4\pi G m_l n_{l0})^{1/2}$ is the Jeans frequency for the LNS. Normalized gravitational potential is given as $\Psi = \psi/C_l^2$. $\Phi = \phi e/m_e c^2$ gives the normalized electrostatic potential.

## 3. Perturbation analysis

The relevant physical fluid parameters ($F$) for the plasma fluid are linearly perturbed ($F_1$) about their hydrostatic homogeneous equilibrium values ($F_0$) using a normal spherical mode analysis [40] in an auto-normalized Fourier form given as

$$F(R,\tau) = F_0 + F_1(R,\tau) = = F_0 + F_{10}\left(\frac{1}{R}\right)\exp\left[-i(\Omega\tau - k^* R)\right], \tag{21}$$

$$F = [N_s \quad M_s \quad \Phi \quad \Psi]^T, \tag{22}$$

$$F_0 = [1 \quad 0 \quad 0 \quad 0]^T, \tag{23}$$

$$F_1 = [N_{s1} \quad M_{s1} \quad \Phi_1 \quad \Psi_1]^T. \tag{24}$$

The spatial and temporal operators get modified in the defined Fourier space $(\Omega, k^*)$ as $\partial/\partial R \to (ik^* - 1/R)$ and $\partial/\partial \tau \to (-i\Omega)$. Here, $\Omega \ (=\omega/\omega_{pl})$ denotes the normalized fluctuation frequency and $k^* \ (=k/2\pi\lambda_{Dl}^{-1})$ designates the normalized wavenumber. The relevant fluid parameters appearing in Equations (13)-(20) in the new wave-space can be written as

$$N_{e1} = -i\Omega^{-1}(ik^* + R^{-1})M_{e1}, \tag{25}$$

$$M_{e1} = i\Omega\Phi_1\{(ik^* + R^{-1})(P_e^* - 4^{-1}H'^2 M_{Fe}^2)\}, \tag{26}$$

$$N_{l1} = -i\Omega^{-1}(ik^* + R^{-1})M_{l1}, \tag{27}$$

$$M_{l1} = \{(ik^* + R^{-1})\Phi_1 - i\sigma\mu'(\Omega\beta k^{*2})^{-1}(k^{*2} + R^{-2})M_{h1}\}L^{-1}, \tag{28}$$

$$N_{h1} = -i\Omega^{-1}(ik^* + R^{-1})M_{h1}, \tag{29}$$

$$M_{h1} = -(ik^* - R^{-1})\{\beta - i\sigma(k^{*2} + R^{-2})(\Omega k^{*2} L)^{-1}\}\Phi_1 H^{-1}, \tag{30}$$

$$\Phi_1 = -k^{*-2}\{(1+\mu')N_{e1} - N_{l1} - \mu' N_{h1}\}, \tag{31}$$

$$\Psi_1 = -\sigma k^{*-2}(N_{l1} + \mu'\beta^{-1}N_{h1}). \tag{32}$$

In the above set of equations, the various substituted terms in an expanded form are given as

$$P_e^* = p_{dp}^* - p_{ie}^* - p_{exc}^* - p_{corr}^*, \quad \text{(for the CD case)} \tag{33.1}$$

$$P_e^* = G_e' T^* - p_{ie}^* - p_{exc}^* - p_{corr}^*, \quad \text{(for the ND case)} \tag{33.2}$$

$$L = i\{\Omega - \Omega^{-1}(k^{*2} + R^{-2})(A_{el}T^* - \sigma k^{*-2})\}, \tag{34}$$

$$H = -i\Omega + (k^{*2} + R^{-2})\left[-i\Omega^{-1}\{\sigma\mu'(\beta k^{*2})^{-1} - A_{eh}T^*\} - \sigma^2(k^{*2} + R^{-2})(\Omega^2 k^{*2}\beta L)^{-1}\right]. \tag{35}$$

A standard procedure of algebraic elimination and simplification among Equations (25)-(32) yields a generalized linear (sextic) dispersion relation on the electrodynamic spatiotemporal response scales of the constitutive LNS with all the generic notations [36, 38] given as

$$\Omega^6 + A_4\Omega^4 + A_2\Omega^2 + A_0 = 0. \tag{36}$$

The different coefficients appearing in Equation (36) can be written in an expanded form as

$$A_4 = (k^{*2} + R^{-2})[-Ek^{*2}T^*(2A_{el} + A_{eh}) + E\sigma(2 + \mu'\beta^{-1}) + (1+\mu')\{-2A_{el}T^* + \sigma k^{*-2}(2+\mu'\beta^{-1})$$
$$- A_{eh}T^* - E\}][Ek^{*2} + (1+\mu')]^{-1}, \tag{37}$$

$$A_2 = (k^{*2} + R^{-2})^2[2Ek^{*2}A_{el}A_{eh}T^{*2} - 2E\sigma\{A_{eh}T^* + (1+\mu'\beta^{-1})A_{el}T^*\} + E\sigma^2 k^{*-2}(1+\mu'\beta^{-1})$$
$$+ Ek^{*2}(A_{el}T^*)^2 - 2\sigma k^{*-2}(1+\mu')\{(1+\mu'\beta^{-1})A_{el}T^* + A_{eh}T^*\} + (1+\mu')A_{el}T^{*2}(2A_{eh} + A_{el})$$
$$+ (\sigma k^{*-2})^2(1+\mu')(1+\mu'\beta^{-1}) + ET^*(A_{el} + A_{eh}) - \sigma k^{*-2}(E+\mu'\beta^{-1}) + 2\mu'\beta E(A_{el} - \sigma k^{*-2})$$
$$+ 2E\mu'\sigma k^{*-2}][Ek^{*2} + (1+\mu')]^{-1}, \tag{38}$$

$$A_0 = (k^{*2} + R^{-2})^3[EA_{el}A_{eh}T^{*2}(2\sigma - A_{el}T^*k^{*2}) - \sigma^2 k^{*-2}T^*(A_{eh} + \mu'\beta^{-1}A_{el})\{E + \sigma^2 k^{*-2}(1+\mu')\}$$
$$+ E\sigma\mu'\beta^{-1}(A_{el}T^*)^2 - (1+\mu')A_{el}^2 A_{eh}T^{*3} - \sigma\mu'(1+\mu')(A_{el}T^*)^3(\beta k^{*2})^{-1} + E\sigma\mu'(\beta k^{*2})^{-1}(1-2\sigma k^{*-2})$$

$$+ EA_{eh}T^*\left(\sigma k^{*-2} - A_{el}T^*\right) + \mu'\left(\sigma k^{*-2}\right)^2\left(2E - \beta^{-1}\right) + \mu'\beta E\left\{\left(A_{el}T^*\right)^2 + \left(\sigma k^{*-2}\right)^2\right\}$$

$$-2E\mu'\sigma k^{*-2} A_{el}T^*(1+\beta)\Big]\Big[Ek^{*2} + (1+\mu')\Big]^{-1}. \tag{39}$$

In Equations (37)-(39), $E = P_e^* - 4^{-1}H'^2 M_{Fe}^2 k^*$. (40)

The dispersion relation given by Equation (36) gets differently modified in the low and the high-frequency limits. However, the high-frequency limit is not discussed here. In the low-frequency limit ($\Omega^q = 0 \, \forall q > 2$), Equation (36) can be written as

$$A_2\Omega^2 + A_0 = 0 . \tag{41}$$

It is worth mentioning here that the mathematical expressions for $A_2 = A_2(R,k^*)$ and $A_0 = A_0(R,k^*)$ are already given by Equations (38)-(39), respectively. A number of interesting WD features is clearly evident from the reduced dispersion relation (Equation (41)) in the low-frequency limit. We use $\Omega = \Omega_r + i\Omega_i$ to characterize the NAW instability behaviours. Here, $\Omega_r$ characterizes the propagatory aspects (where, $v_p = \Omega_r/k^*$, $v_g = d\Omega_r/dk^*$). In contrast, $\Omega_i$ depicts the growth/damping behaviour of the same by making the wave amplitude modulated by a factor of $\exp(\pm\Omega_i\tau)$ [41]. The dynamics of the NAWs in the CD states of the ONe and CO cores and their ND transition regions is significantly influenced conjointly by the concentrations of the constituent species, their mutual electrostatic interactions, temperatures, masses, and so forth.

## 4. Results and discussions

The stability, propagatory, dispersive nature of the NAWs excited in the CD ONe and CO WD cores and the ND transition region around the cores are analyzed herein using a three-component quantum hydrodynamic plasma model. The quantum electronic species evolves under the conjoint pressures due to temperature degeneracy, interaction with surrounding electrons and other nuclei, exchange energy, and correlation energy. Likewise, the classical thermal pressures are retained for the larger classical species (LNS+HNS). A normal spherical mode analysis over the considered system yields a generalized linear sextic dispersion relation (Equation (36)), which is modified using the low-frequency approximation (Equation (41)). A numerical illustrative platform is provided to reveal the nature of the derived dispersion relation in the low-frequency regime (Equation (41)). The growth rate corresponding to the NAW instability, its propagatory and dispersive features are illustrated pictorially in Figures 1-20. To get a clear idea of the dispersive nature, we use illustrative Matlab plots depicting the phase dispersion and group dispersion, in addition to the phase and group velocities. The different input values used herein have been calculated using preliminary data available in different trustworthy and reliable literary sources [3, 26-29, 32, 34, 42].

*4.1 Analysis of the CD ONe core*

In Figure 1, we depict the profile structures of the normalized real angular frequency ($\Omega_r$) (Figure 1(a)) and normalized imaginary angular frequency ($\Omega_i$) (Figure 1(b)) with the normalized angular wavenumber ($k^*$) for different values of the thermodynamic temperature ($T$). The different coloured lines link to $\Omega_r$ for $T = 6 \times 10^9$ K (blue solid line), $T = 7 \times 10^9$ K (red dashed line), and $T = 8 \times 10^9$ K (black dotted line). Figure 1 clearly shows that $\Omega_i$ exists in the low $k^*$-space (Figure 1(b)), indicating an unstable behaviour. For a given value of $T$, $\Omega_i$ increases with increasing $k^*$, becomes maximum, and then decreases to zero. For gradually increasing values of $T$, the $k^*$-value at which $\Omega_i$ attains peak shifts towards the left side of the $k^*$-axis, that is, towards smaller $k^*$-values. As we proceed towards higher values of $k^*$, we have $\Omega_r$ (Figure 1(a)), indicating the propagatory nature of NAWs.

In Figure 2, we depict the profile of the phase velocity ($v_p$) in the same conditions as Figure 1. The different coloured lines link to different $v_p$ for $T = 6 \times 10^9$ K (blue solid line), $T = 7 \times 10^9$ K (red dashed line), and $T = 8 \times 10^9$ K (black dotted line). It is clearly seen that for a given value of $T$, $v_p$ increases with $k^*$. Thus, $v_p$ depends on $k^*$, indicating the dispersive nature of the system [41, 43-45]. $v_p$ gives the speed of travelling waves. In other words, $v_p$ denoted by the blue solid line indicates the velocity of the NAW at $T = 6 \times 10^9$ K, and so on. Higher the $T$ of the core, higher is the $v_p$, and vice-versa.

Figure 3 depicts the group velocity ($v_g$) profile with variation in $k^*$ for different indicated values of $T$. Different coloured lines correspond to different $v_g$ for $T = 6 \times 10^9$ K (blue solid line), $T = 7 \times 10^9$ K (red dashed line), and $T = 8 \times 10^9$ K (black dotted line). Considering the clearly visible trends depicted by Figure 3(a), it can be fairly commented that for a given value of $T$, $v_g$ first forms a peak, then decreases and becomes almost constant with increasing $k^*$. As the value of $T$ increases, $v_g$-peak increases and vice-versa (Figure 3(b)). It is a well-known fact that $v_g$ is the velocity at which a bump travels in a wave [45]. In general, the macroscopic NAW propagates through the plasma medium consisting of spectral components of many different acoustic frequencies. If these components are to travel together, then they form a bump as per the wave packet model. A bump is essentially the point at which the phases of the different components become equal and thus, add constructively forming a peak. However, due to different speeds and phases of the different components, the peak gradually dissolves. A second peak may be observed when the phase and amplitude coordinations among the different components take place [43-45].

Figure 4 gives the profile of the phase dispersion ($D_p = \partial_{k^*} v_p$) [41, 44] in the same conditions as Figure 1. As clearly indicated, the different lines connect to different $D_p$ for $T = 6 \times 10^9$ K (blue solid line), $T = 7 \times 10^9$ K (red dashed line), and $T = 8 \times 10^9$ K (black dotted line). Gradually increasing $T$ leads to gradual $D_p$ enhancement. However, the $k^*$-value at which $D_p$ attains peak (maxima) shifts towards left with increase of $T$ (as clearly indicated by Figure 4(b)). Figure 4(c) depicts the magnified version in the range $k^* = 4.2 - 5$. The dispersive nature of the system in terms of the NAW response is further confirmed in Figure 4.

As in Figure 5, we depict the group dispersion ($D_g = \partial_{k^*} v_g$) [41, 44] in the same conditions as Figure 1. The different lines link to different $D_g$ for $T = 6 \times 10^9$ K (blue solid line), $T = 7 \times 10^9$ K (red dashed line), and $T = 8 \times 10^9$ K (black dotted line). As we proceed from $k^* = 1$ towards $k^* = 1.3$, we observe that $D_g$ lines tend to decrease, that is frequency (angular wavenumber) shift and $D_g$ follow opposite trends, implying an unstable situation [44]. However, on moving towards right starting from $k^* = 1.3$, $D_g$ starts to increase again with increase of $k^*$, implying a propagatory nature. This further confirms the obtained $k^*$-range (Figure 1) for $\Omega$ to show unstable ($\Omega_i$) and propagatory ($\Omega_r$) behaviours.

## *4.2 Analysis of the ND transition region around ONe core*

Figure 6 shows the same as Figure 1, but for the ND transition region around the ONe core. The colour coding of the three lines is the same as Figures 1-5. As in Figure 1, we find that $\Omega_i$ exists for lower $k^*$-values (Figure 6(b)). In Figure 6(b), for a particular $T$, $\Omega_i$ first increases, becomes maximum and then decreases back. As $T$ gradually increases, the peak at which $\Omega_i$ becomes maximum shifts towards the left side in the $k^*$-axis. As we move towards right in the $k^*$-axis, that is, as the value of $k^*$ increases, we have real values of $\Omega$, that is, $\Omega_r$ (Figure 7(a)), which indicates propagatory behaviour of the NAW in the high-$k^*$ space. A spike is observed in both the Figures 6(a)-6(b), unlike previous Figures 1(a)-1(b).

Figure 7 shows the same as Figure 2, but for the borderline region around the core. The different coloured lines indicate that $v_p$ increases with increasing $k^*$, thereby showing that $v_p$ depends on $k^*$. Thus, the system is dispersive. It is also observed that $v_p$ increases with increasing $T$. In other words, higher $T$ indicates higher $v_p$, same as the CD core (Figure 2).

Figure 8 shows the same as Figure 3, but for the surrounding ND borderline region depicting the transition between $T$ and $T_F$. It is clearly seen from Figure 8(a) that $v_g$ forms a peak and then decreases and becomes almost constant with gradually increasing $k^*$. However, as $T$ gradually increases, $v_g$ decreases (Figure 8(b)), in contrast to the CD ONe core (Figure 3). Thus, the point at which the phases of the different components of the wave add up constructively to form a peak is lower for higher $T$. Thus, in the ND borderline region, $v_g$ decreases with increasing $T$.

In a similar fashion, Figure 9 shows the same as Figure 4, but for the ND region around the core. The $D_p$ profiles further confirm the dispersive nature of the plasma. It is seen that the features exhibited by the $D_p$ curves in the ND surrounding of the core are the same as that in CD core (Figure 4). $D_p$ increases with increasing $T$ and the $D_p$ maxima (peak) shifts towards the left of the $k^*$-axis with $T$ enhancement. Figures 9(b)-9(c) show the sectional magnified versions of Figure 9(a) in the range $k^* = 0.9 - 2$ and $k^* = 4 - 4.8$, respectively.

Likewise, Figure 10 shows the $D_g$ profile with variation in $k^*$ for different $T$ values, that is, same as Figure 5, but in the ND region. If we proceed from $k^* = 1.2$ towards $k^* = 1.5$, we see that $D_g$ decreases. Thus, $D_g$ and $k^*$ follow opposite trends ($D_g$ decreases with increase of $k^*$), thereby indicating unstable behaviour in the low-$k^*$ space [44]. However, as we proceed from $k^* = 1.5$ towards right in the $k^*$-axis, $D_g$ increases. Thus, $D_g$ and $k^*$ follow the same trend, showing a propagatory behaviour in the high-$k^*$ space. This reinforces the accuracy of the trends obtained in Figures 6(a)-6(b). In other words, $\varOmega_r$ (propagatory NAW) exists from $k^* = 1.5$ onwards (Figure 6(a)), whereas $\varOmega_i$ (growing NAW) exists in the low-$k^*$ space till $k^* = 1.5$ (Figure 6(b)).

## 4.3 Analysis of the CD CO core

Figure 11 shows the same as Figure 1, but for the CD CO cores. The different coloured lines link to different $\varOmega_r$ (Figure 11(a)) and $\varOmega_i$ (Figure 11(b)) values with variation in $k^*$ for $T = 2 \times 10^6$ K (blue solid line), $T = 2 \times 10^7$ K (red dashed line), and $T = 2 \times 10^8$ K (black dotted line). The observations are the same as the CD ONe core (Figures 1(a) and 1(b)), that is, for a given $T$, $\varOmega_i$ (unstable) exists for the low-$k^*$ values (Figure 11(b)). As we move towards high-$k^*$ values, we get $\varOmega_r$, which implies propagatory behaviour (Figure 11(a)). The only difference observed between Figure 1 and Figure 11 is the $k^*$-range in which $\varOmega$ shows unstable and propagatory behaviours. The $k^*$-range in Figure 11 is observed to be much higher than that in Figure 1.

Figure 12 shows the same as Figure 2, but for the CD core. The various coloured lines correspond to different $v_p$ for $T = 2 \times 10^6$ K (blue solid line), $T = 2 \times 10^7$ K (red dashed line), and $T = 2 \times 10^8$ K (black dotted line). For a given $T$, $v_p$ starts increasing from the $k^*$-point from which $\varOmega_r$ comes into existence, after which the $v_p$-curve attains an almost constant value with increasing $k^*$. That is, $v_p$ gives the velocity of the propagatory NAW. For increasing $T$, $v_p$ increases and vice-versa. It implies that $v_p$ is $k^*$-dependent and hence, the system is dispersive [44, 45].

Figure 13 shows the same as Figure 3, but for the CO core. The different lines link to different $v_g$ for $T = 2 \times 10^6$ K (blue solid line), $T = 2 \times 10^7$ K (red dashed line), and $T = 2 \times 10^8$ K (black dotted line). For a given $T$, $v_g$ starts increasing from the $k^*$-point from which $\varOmega_r$ comes into existence, attains peak for a very small $k^*$-range, after which it starts to decrease again. $v_g$-curve attains an almost constant value with increasing $k^*$ after decreasing from the peak. The peak attained is highest for $T = 2 \times 10^8$ K, followed by $T = 2 \times 10^7$ K and so on. With increase of $T$, the peaks attained by the $v_g$-curves shift towards the smaller $k^*$-values. It clearly shows that the phases of the different components of the NAW add up constructively for a very short frequency range, thus forming the $v_g$-peaks for very small $k^*$-range.

The dispersive nature of the considered plasma system is further confirmed by Figure 14, which shows the same as Figure 4, but for the CD CO WD core. As clearly indicated by Figures 14(a)-14(b), the different coloured lines link to different $D_p$ for $T = 2\times 10^6$ K (blue solid line), $T = 2\times 10^7$ K (red dashed line), and $T = 2\times 10^8$ K (black dotted line). The observations are the same as Figure 4, except for the fact that the $D_p$ exists for a much higher $k^*$-range than that in Figure 4. $D_p$ first increases from the $k^*$-value from which $\Omega_r$ comes into existence. It then attains peak and starts to decrease towards very low $D_p$ value. With increase of T, the $D_p$-peak shifts towards low-$k^*$ values and vice-versa (as clearly seen from Figure 14(a)).

Figure 15 shows the same as Figure 5, but for the CD CO core. The different lines correspond to different $D_g$ for $T = 2\times 10^6$ K (blue solid line), $T = 2\times 10^7$ K (red dashed line), and $T = 2\times 10^8$ K (black dotted line). In the low-$k^*$ space, $D_g$ forms a peak and then tends to decrease with increasing $k^*$. Thus, $D_g$ and $k^*$ follow opposite trends, indicating an unstable situation. However, starting from the point at which different $\Omega_r$ for different T come into existence, the $D_g$-curves start to increase again with increase of $k^*$, thus showing propagatory nature of the considered NAW. This further re-confirms the obtained $k^*$-range for existence of $\Omega_r$ and $\Omega_i$ (Figures 11(a)-11(b)).

*4.4 Analysis of the ND transition region around CO core*

Figure 16 shows the same as Figure 6, but for the ND transition region around the CO core. The colour coding adopted here is the same as Figures 11-15. It is observed that for the low-$k^*$ space, $\Omega_i$ exists for different T (Figure 16(b)), thereby indicating an unstable behaviour. The $\Omega_i$-peaks shift towards the smaller $k^*$-values with increasing T. As we move towards higher $k^*$-values, $\Omega_r$ comes into existence (Figure 16(a)), indicating the propagatory nature of NAW.

Figure 17 shows the same as Figure 7, but for the ND region around the core. It can be clearly observed that $v_p$ depends on $k^*$ for a given T. This clearly hints towards the dispersive nature of the considered plasma system. In addition, $v_p$ increases with increasing T.

Figure 18 shows the same as Figure 8, but for the ND transition region around the CO core. The peak at which the phases of the different components of the propagatory NAW become equal and add up constructively exists for a very small $k^*$-range for a particular T. With gradually increasing T, the $v_g$-peak shifts towards the low $k^*$-values.

Figure 19 depicts the same as Figure 9, but for the ND region around the CO core. The observations are the same as Figure 9, except for the fact that $D_p$ exists for a much higher $k^*$-range as compared to Figure 9. It confirms the dispersive nature of the plasma medium, which is the same as Figure 9. The $D_p$-peaks shift towards the lower $k^*$-values with T enhancement.

Figure 20 shows the same as Figure 10, but for the borderline region around the CD CO core. In the low-$k^*$ space, $D_g$ forms a peak and then tends to decrease with increasing $k^*$. Thus, $D_g$ and $k^*$ follow opposite trends, indicating an unstable situation. However, starting from the point at which different $\Omega_r$ for different $T$ come into existence, the $D_g$-curves start to increase again with increase of $k^*$, thus showing propagatory nature of the considered NAW. This further affirms the obtained $k^*$-range for the existence of $\Omega_r$ and $\Omega_i$ (Figures 16(a)-16(b)).

## 5. Conclusions

A three-component spherically symmetric plasma model is developed to analyze the stability, propagatory, and dispersive behaviour of the NAWs excited in the CD ONe WD cores, CO WD cores, and their ND surroundings. The model comprises of quantum electrons and classical LNS-HNS initially in a hydrostatic homogeneous equilibrium configuration. The constitutive electrons are acted upon by the pressures due to temperature degeneracy, interaction with surrounding electrons and other nuclei, exchange energy, and correlation energy explicitly. The pressures due to exchange and correlation are purely quantum-mechanical in origin and therefore have no classical analogs. The classical thermal pressure acting upon LNS and HNS are retained in their respective EoSs. A standard normal spherical mode analysis yields a sextic generalized linear dispersion relation, for low-frequency fluctuation analysis. It is seen that for both the CD core and the ND transition region around the core, we get propagatory NAW ($\Omega_r$) as we move towards high-$k^*$ space (Figures 1(a), 6(a), 11(a), 16(a)). Unstable behaviour ($\Omega_i$) exists for the low-$k^*$ space (Figures 1(b), 6(b), 11(b), 16(b)). The NAW $v_p$ show similar trends in both the CD core and the ND region around the core for both the dwarfs (Figures 2, 7, 12, 17). In both the cases, $v_p$ is dependent on $k^*$, thereby indicating dispersive nature of the considered plasma medium. In the case of the CD ONe core, the point at which different components of the propagatory NAWs add up constructively to form a peak gradually increases with increase of thermodynamic temperature (Figure 3), in contrast to ND region around the core (Figure 8). However, both the CD CO core and ND transition region around the core show a common trend, that is, higher $v_g$-peaks with increase of $T$ (Figure 13, 18). $D_p$ observed for the CD core and ND region around the core of both the ONe and CO WDs follow the same trend, thus re-confirming the dispersive nature of the medium (Figures 14, 19). The trends observed for $D_g$ for both the ONe and CO CD cores and ND transition region around the cores reaffirm the observed $k^*$-ranges for the propagatory and unstable behaviour of the proposed NAW modes (Figures 15, 20). Besides, an appendix highlighting a clear tabular distinction between the two considered WDs is added at the end. It may be noteworthy that, unlike the presented pulsational study (radial), several pulsations (angular) have previously been reported in PG1159 pre-WDs, variable DB, and variable DA [2]. It implicates that there are fair possibilities for the detection of the proposed modes in dwarf family stars and closely related compact astrophysical circumstances in the near future with the needful refinements in modern astronomy and space exploration systems [46].

At the last, we are strongly hopeful that our semi-analytic and semi-classical analysis presented here may open a hotspot area of emerging research in the context of diversified collective acoustic waves, oscillations, and instabilities excitable in the ONe and CO WDs and similar compact astrophysical circumstances in an important asteroseismic investigative direction [45-47].

**Acknowledgements.** Active cooperation received from Tezpur University is thankfully acknowledged. The financial support received through the SERB Project (Grant- EMR / 2017 / 003222) is duly recognized.

**Disclosure statement**
No potential conflict of interest was reported by the authors.

# Appendix: Comparison between ONe WD and CO WD

| S. No. | Item | ONe WD | CO WD |
|---|---|---|---|
| 1. | Progenitor mass | $8M_\Theta \lesssim M \lesssim 11 M_\Theta$ [7] | $M \leq 8 M_\Theta$ [4] |
| 2. | WD core mass | $(1.2 - 1.37) M_\Theta$ [32, 42] | $0.6 M_\Theta$ [34] |
| 3. | Unstable behaviour zone | $k^* = 0-1.2$ to $k^* = 0-1.4$ in CD case (Figure 1(b)); $k^* = 0-1.13$ to $k^* = 0-1.22$ in ND case (Figure 6(b)); Spikes found in ND case in $k^* = 1.24 - 1.37$ | $k^* = 0-7.8$ to $k^* = 0-77.3$ in CD case (Figure 11(b)); $k^* = 0-8.1$ to $k^* = 0-76.9$ in ND case (Figure 16(b)) No spike observed in any case |
| 4. | Propagatory behaviour | $k^* = 1.1-1.3$ onwards in CD case (Figure 1(a)); $k^* = 1.27-1.36$ onwards in ND case (Figure 6(a)); Spikes exist in ND case in $k^* = 1.12 - 1.25$ | $k^* = 7.7-77.2$ onwards in CD case (Figure 11(a)); $k^* = 7.2-76.8$ onwards in ND case (Figure 16(a)) |
| 5. | Temperature sensitivity of NAW parameters | More $T$-sensitive (Figures 1-10) | Less $T$-sensitive (Figures 11-20) |
| 6. | Phase velocity of NAW propagation | Increases with both $k^*$ and $T$ (Figures 2, 7) | Same as ONe core case (Figures 12, 17) |
| 7. | Group velocity of NAW propagation | Increases with $T$ in CD case (Figure 3); Decreases $T$ in the ND case (Figure 8) | Increases with $T$ in both the CD and ND cases (Figures 13, 18) |
| 8. | Phase dispersion | Sensitively depends on $k^*$ in both CD and ND cases | Same as ONe core case (Figures 14, 19) |

| | | (Figures 4, 9) | |
| --- | --- | --- | --- |
| 9. | Group dispersion | Prominently varies in the spectral zone $k^* = 0.8 - 1.8$ and peaks shift towards lower-$k^*$ region with higher $T$ (Figures 5, 10) | Prominently varies in the spectral zone $k^* = 7 - 90$ and peaks shift towards lower-$k^*$ region with higher $T$ (Figures 15, 20) |

**Figures**

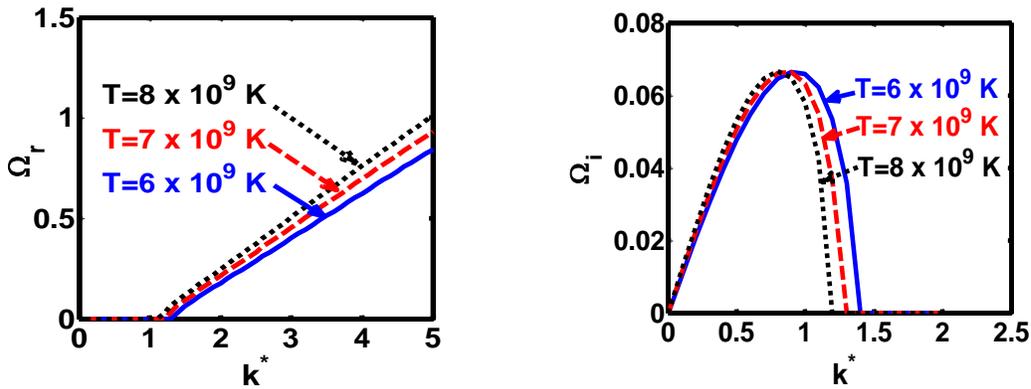

Figure 1. Profile of the normalized (a) real angular frequency ($\Omega_r$) and (b) imaginary angular frequency ($\Omega_i$) with the normalized angular wavenumber ($k^*$) for different values of the thermodynamic temperature ($T$) in the completely degenerate (CD) case of the ONe WD core. The standard normalization scheme with respect to the electrodynamic spatiotemporal response scales of the constitutive LNS in the context of compact astroplasmas is described in the text.\

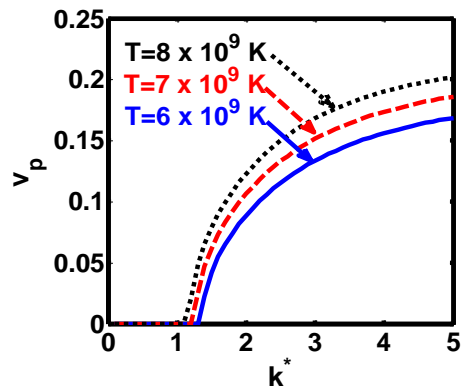

Figure 2. Profile of the normalized NAW phase velocity $(v_p)$ in the same conditions as Figure 1.

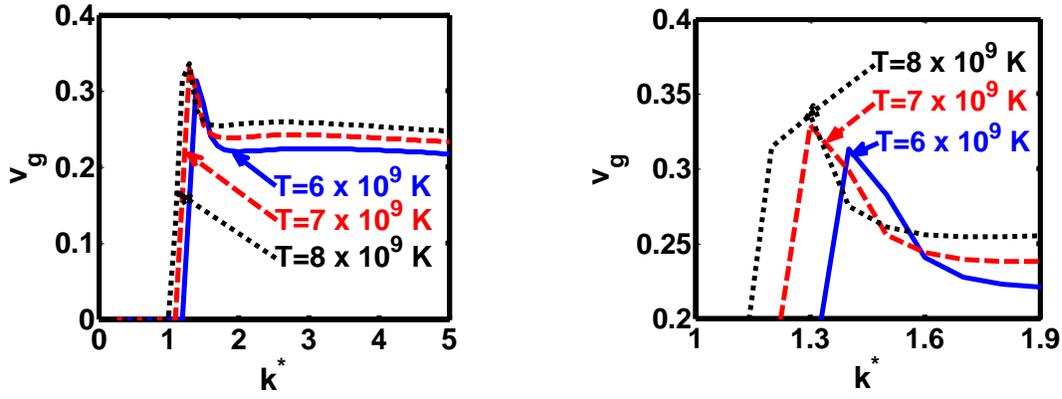

Figure 3. Profile of the normalized NAW group velocity $(v_g)$ of the NAWs in the same conditions as Figure 1. The distinct panels depict $v_g$ in: (a) $k^* = 0-5$ and (b) $k^* = 1-1.9$.

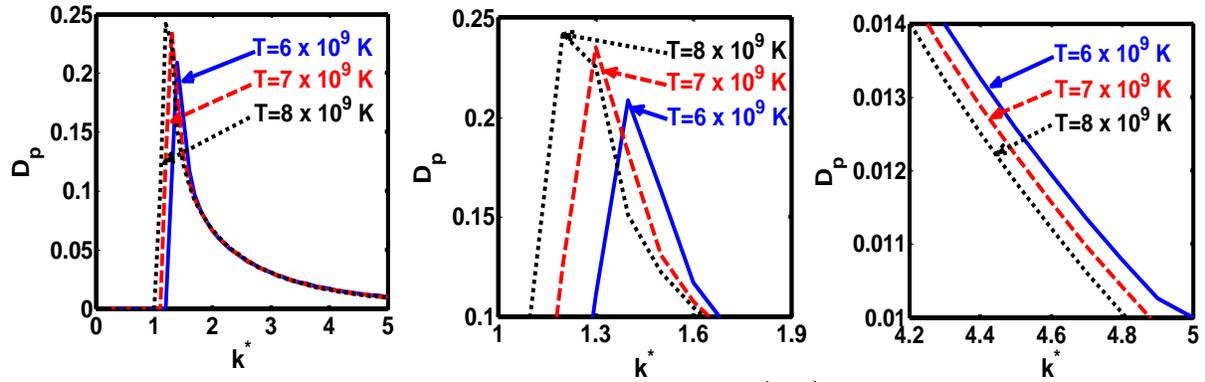

Figure 4. Profile of the normalized NAW phase dispersion $(D_p)$ in the same conditions as Figure 1. The distinct panels give $D_p$ in: (a) $k^* = 0-5$, (b) $k^* = 1-1.9$, and (c) $k^* = 4.2-5$.

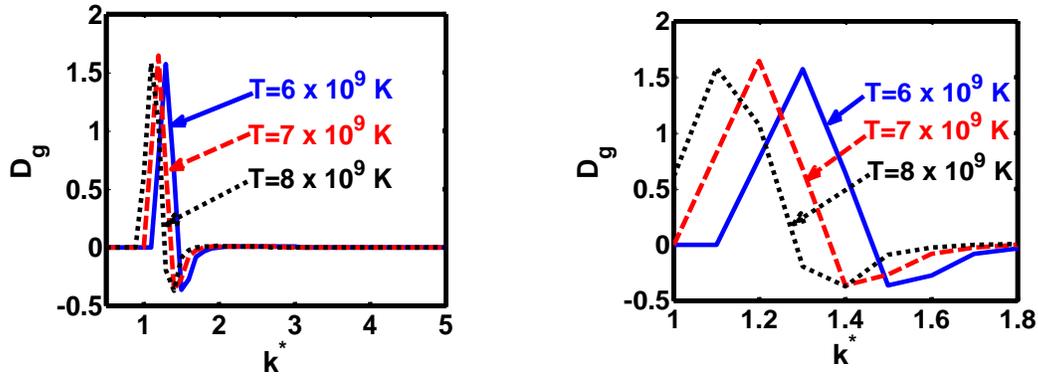

Figure 5. Profile of the normalized NAW group dispersion $(D_g)$ in the same conditions as Figure 1. The distinct panels give $D_g$ in: (a) $k^* = 0-5$ and (b) $k^* = 1-1.8$.

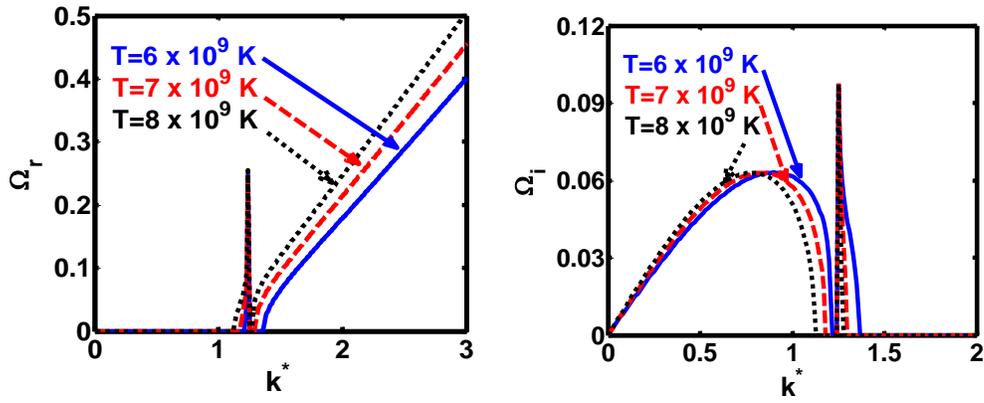

Figure 6. Same as Figure 1, but for the nearly degenerate (ND) case.

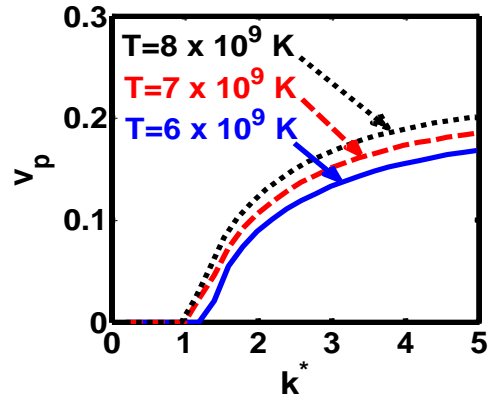

Figure 7. Same as Figure 2, but for the ND case.

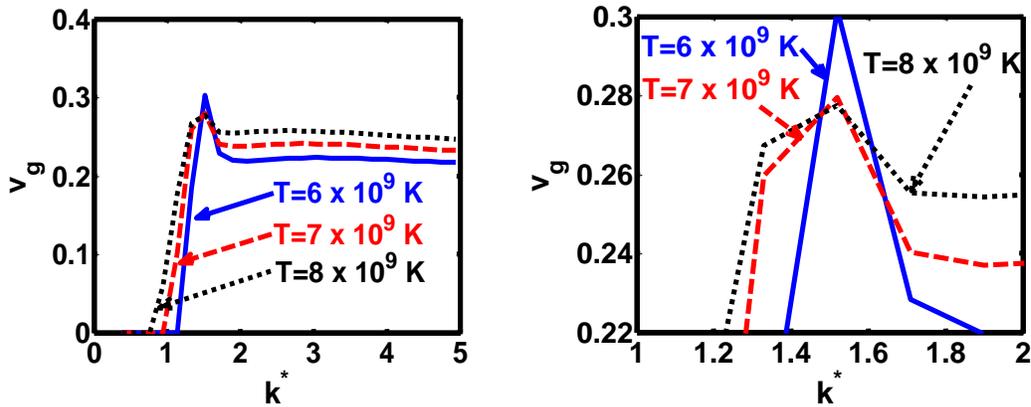

Figure 8. Same as Figure 3, but for the ND case. The distinct panels depict the same in: (a) $k^* = 0-5$ and (b) $k^* = 1-2$.

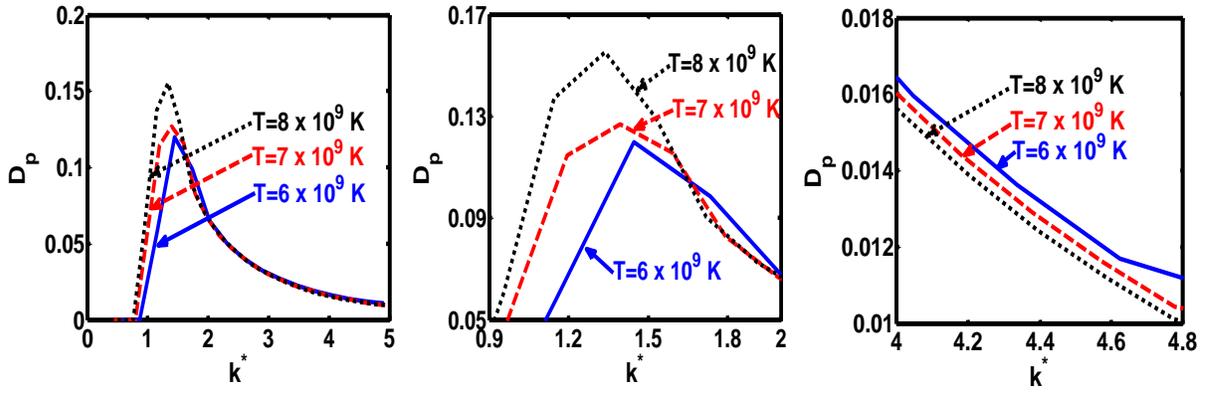

Figure 9. Same as Figure 4, but for the ND case. The distinct panels depict the same in: (a) $k^* = 0 - 5$ (b) $k^* = 0.9 - 2$, and (c) $k^* = 4 - 4.8$.

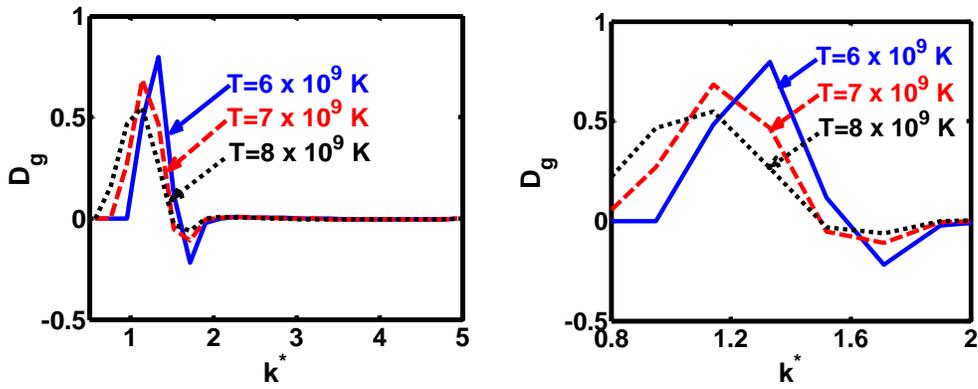

Figure 10. Same as Figure 5, but for the ND case. The distinct panels depict the same in: (a) $k^* = 0 - 5$ (b) $k^* = 1 - 2$.

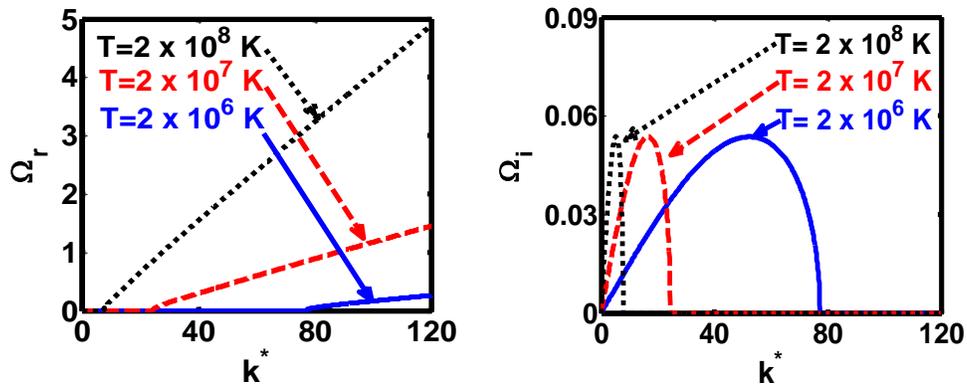

Figure 11. Same as Figure 1, but for CD CO white dwarf (WD) core.

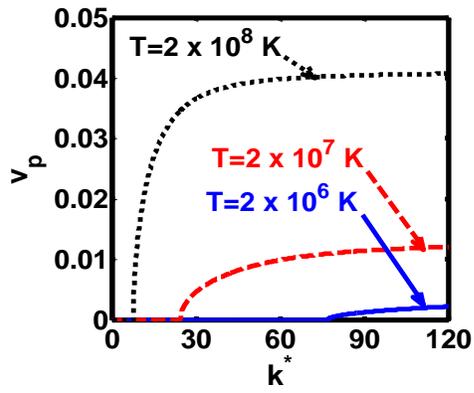

Figure 12. Same as Figure 2, but for CD CO WD core.

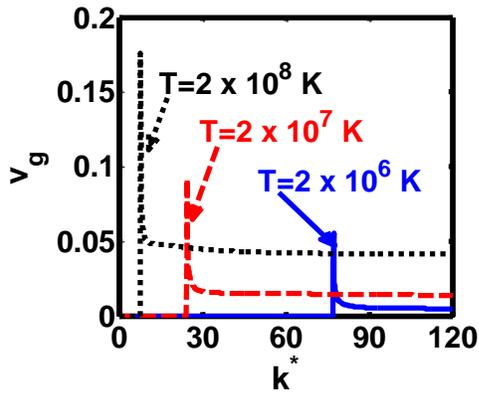

Figure 13. Same as Figure 3, but for CD CO WD core.

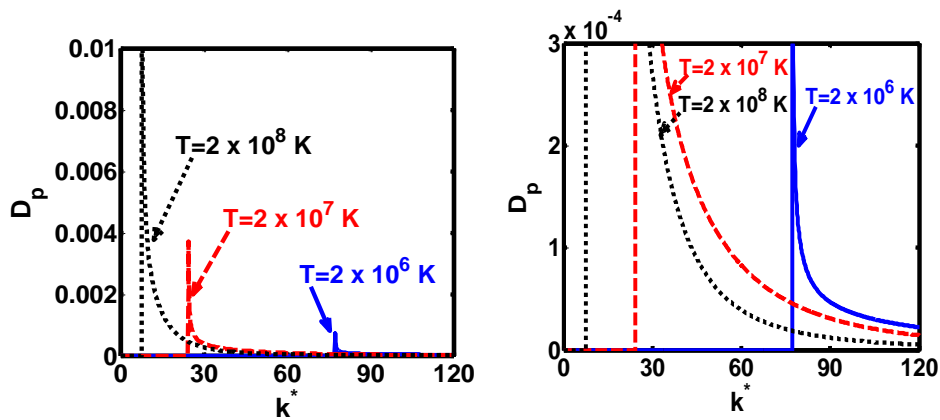

Figure 14. Same as Figure 4, but for CD CO WD core.

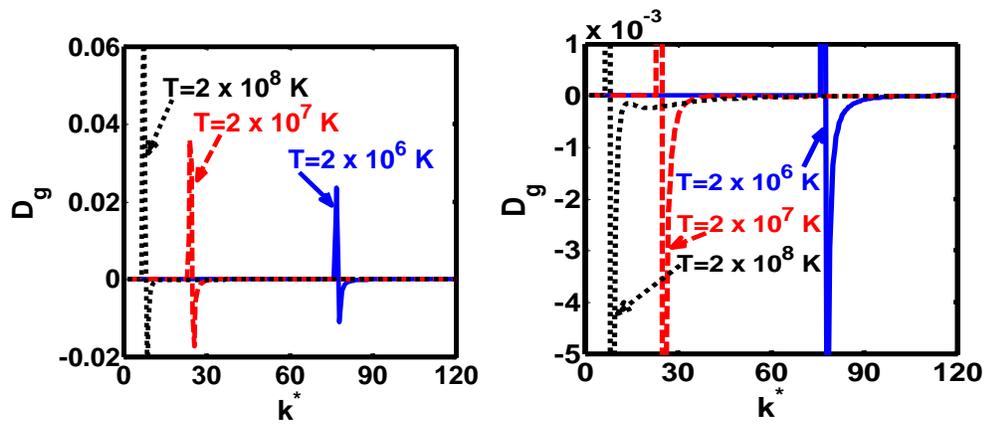

Figure 15. Same as Figure 5, but for CD CO WD core.

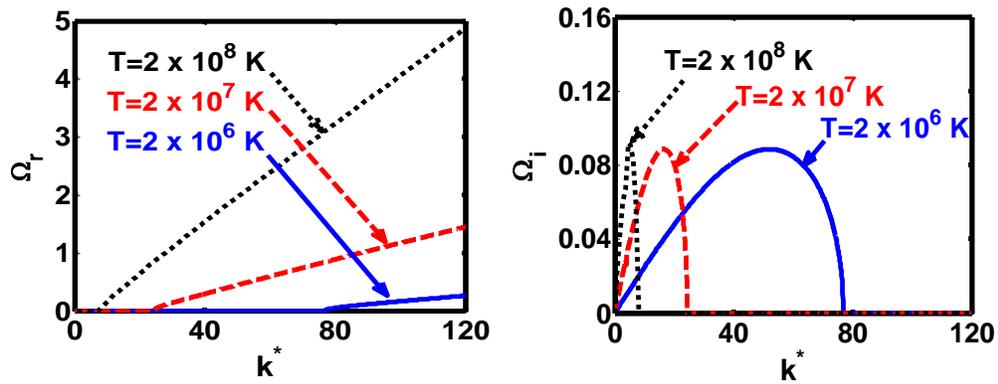

Figure 16. Same as Figure 6, but for CO ND transition region.

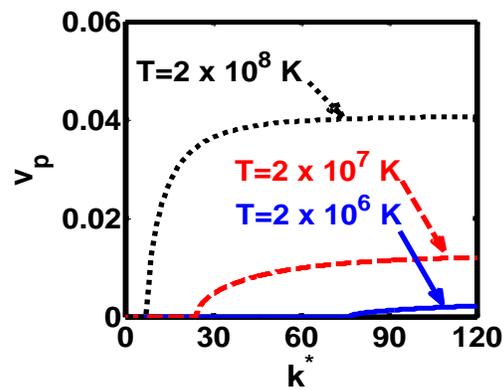

Figure 17. Same as Figure 7, but for CO ND transition region.

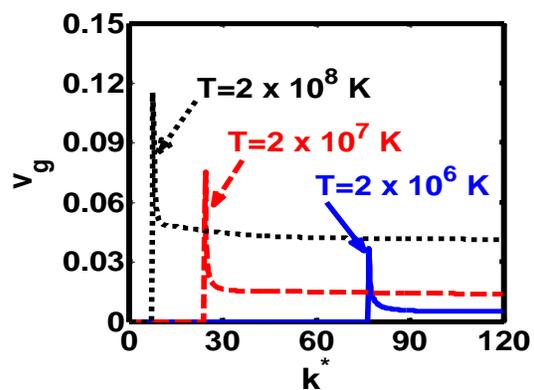

Figure 18. Same as Figure 8, but for CO ND transition region.

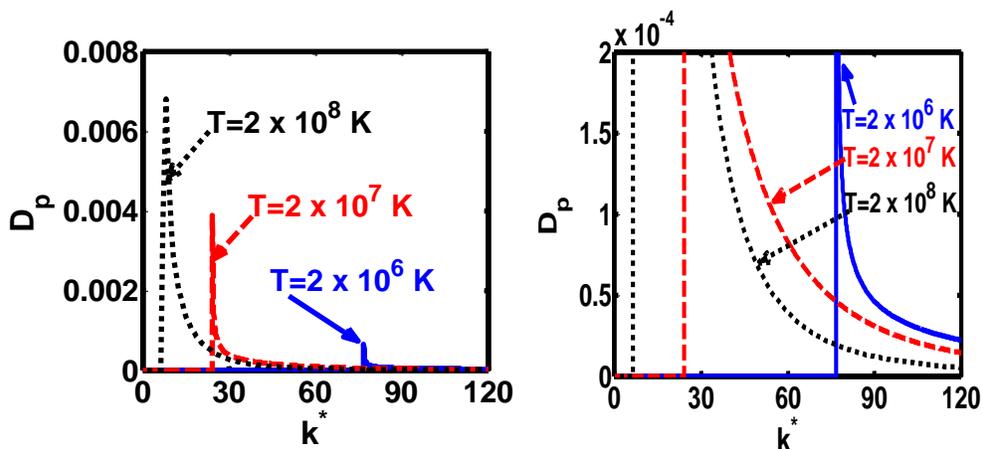

Figure 19. Same as Figure 9, but for CO ND transition region.

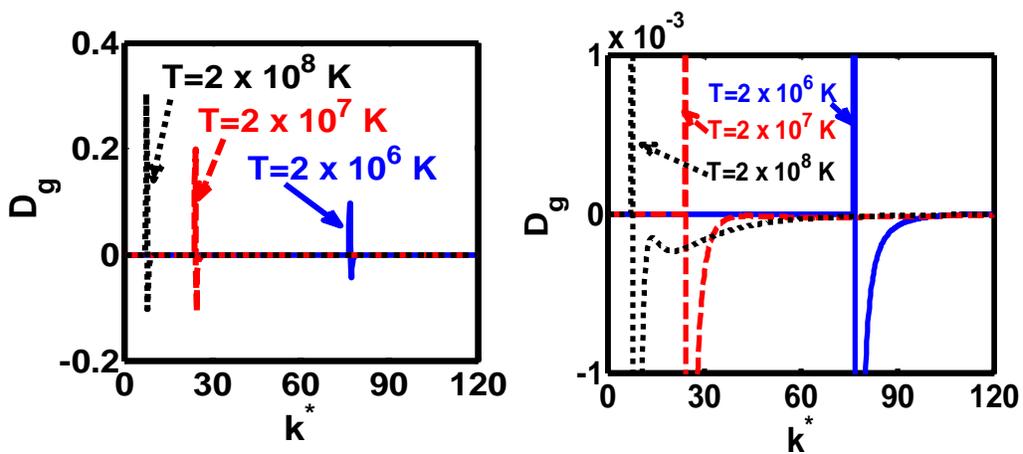

Figure 20. Same as Figure 10, but for CO nearly ND transition region.